\documentclass[a4paper,fleqn]{cas-dc}

\usepackage[numbers]{natbib}

\usepackage{xurl} 
\usepackage{hyperref}
\usepackage{color}
\usepackage{subfigure}
\usepackage{floatrow}
\usepackage{graphicx,epstopdf}
\usepackage{ragged2e}
\usepackage{mciteplus}
\usepackage{chngcntr}
\usepackage[english]{babel}
\usepackage{changepage}

\usepackage{caption} 
\usepackage{threeparttable}
\usepackage[english]{babel}
\usepackage{float}
\usepackage{chngcntr}
\usepackage{enumitem}

\def\tsc#1{\csdef{#1}{\textsc{\lowercase{#1}}\xspace}}
\tsc{WGM}
\tsc{QE}
\tsc{EP}
\tsc{PMS}
\tsc{BEC}
\tsc{DE}

\begin{document}
	\let\WriteBookmarks\relax
	\def\floatpagepagefraction{1}
	\def\textpagefraction{.001}
	\shorttitle{Stationarity of the detrended price return in stock markets}
	\shortauthors{K Arias Calluari et~al.}
	
	\title [mode = title]{Stationarity of the detrended price return in stock markets}                      
	\author[1]{Karina Arias-Calluari}
	\fnmark[1]
	\address[1]{School of Civil Engineering, The University of Sydney, Australia}
	
	\author[2]{Morteza. N. Najafi}
	\address[2]{Department of Physics, University of Mohaghegh Ardabili, Ardabil, Iran}
	
	\author[3]{Michael Harr\'e}
	\address[3]{Complex Systems Research Group, Faculty of Engineering, The University of Sydney, Australia}
	
	\author[1]{Fernando Alonso-Marroquin}
	
	\fntext[fn1]{ https://orcid.org/0000-0001-6013-5490 }
	\nonumnote{karina.ariascalluari@uni.sydney.edu.au}
	
	\begin{abstract}
This paper proposes a governing equation for stock market indexes that accounts for non-stationary effects. This is a linear Fokker-Planck equation (FPE) that describes the time evolution of the probability distribution function (PDF) of the price return. By applying Ito's lemma, this FPE is associated with a stochastic differential equation (SDE) that models the time evolution of the price return in a fashion different from the classical Black-Scholes equation. Both FPE and SDE equations account for a deterministic part or trend, and a stationary, stochastic part as a q-Gaussian noise. The model is validated using the S\&P500 index's data. After removing the trend from the index, we show that the detrended part is stationary by evaluating the Hurst exponent of the multifractal time series,  its power spectrum, and its autocorrelation.\\ 
	\end{abstract}

	\begin{keywords}
		Linear Fokker-Planck equation\sep Stochastic Difussion Equation
		 \sep Hurst exponent \sep Data analysis
	\end{keywords}

	\maketitle

\section{Introduction}
Complex dynamic systems encountered in industry or business \citep{Box2013}, financial markets \citep{nava2016time,gu2010detrending}, genetics \citep{peng1994mosaic}, neuroscience \citep{marton2014detrended,oliveira2019analysis}, biomedicine \citep{huang1998engineering,chua2010application}, ocean dynamics \citep{gay2018stochastic} and seismology \citep{marano2019non} generate non-stationary time series. 
A time series is non-stationarity when it is non-identically distributed throughout their full length of time \citep{politi2012near,mandelbrot2010mis,bahrisensitivity}.
In stock markets, these type of time series are simple price return \citep{nava2016time,bahrisensitivity,maganini2018investigation}, volatility of index price \citep{roman2008skewness} and trading stock volume
\citep{lan2010distribution}. In the statistical analysis of financial markets, the non-stationarity time series that is extensively analyzed is the simple price return, which is an arithmetical difference \citep{lan2010distribution}. For price return, the non-stationarity is produced by the non-constant activity during a trading day and the heterogeneity of market participants
\citep{ponta2019modeling,clara2017diffusive,nava2016time}.

The non-stationary time series are difficult to analyze because the process remains in a non-equilibrium state. Consequently, their models do not achieve optimal forecasting and control \citep{Box2013}. Therefore, it is challenging to obtain an accurate model that can be used to calculate the probability of future value based on initial conditions.\par

In recent years extensive research has been devoted to model the index price return distributions considering its characteristic features and non-stationarity properties \cite{ahn2018modeling,meng2015quantum,meng2016quantum}.  The Black Scholes equation had been the novel governing equation to describe the characteristics of the stock return distribution \cite{chen2016numerically}. This important model assumes the geometric Brownian motion (GBM) equation to model the Index price return. However, it fails to capture the non-Gaussian properties of the price return \cite{chen2016numerically}, negative skewness \cite{ahn2018modeling}, and negative values obtained from recent events \cite{OIL}. As a result, different models started to appear, applying modifications or generalizations to the GBM. The fractional Brownian motion (FBM) \cite{ali2020fractional}, the mixed fractional Brownian motion (MFBM) as a combination of the Brownian motion with the FBM \cite{rao2016pricing}, the generalized mixed fractional Brownian motion (GMFBM) \cite{he2014pricing,thale2009further}. These new models capture better the characteristic features of the Index price return. However, they were developed considering a neutral risk assumption \cite{he2014pricing}. As an alternative to these traditional stock return models, an increasing number of quantum models have also been applied to study the stochastic dynamics of stock prices. The most well-known are; the quantum spatial-periodic harmonic model (QSH) \cite{meng2015quantum}, the quantum Brownian motion model (QBM) \cite{meng2016quantum}, the quantum harmonic oscillator model (QHO) \cite{ahn2018modeling}. An advantage of these models is that they incorporate the market uncertainty (risk), although they have been derived from the stochastic equation of a Brownian process.\par
In this paper, a new governing equation of non-stationary simple price return is presented. This governing equation is based on a simpler but more effective approach. The time-series is decomposed by a deterministic part or trend and a stochastic part or q-gaussian noise. The stochastic part is stationary and can be modeled by a q-gaussian distribution. To validate our governing equation, the trend of the SP\&500 index data is removed by applying a moving average method (MA) based on the Kurtosis evaluation \citep{gu2010detrending}.
By applying different tests, it is shown that the detrended price return is a stationary process. Thus, the PDF of the price return obeys the proposed partial differential equation (PDE).
\begin{figure}[!htbp]
	\centering
	\includegraphics[scale=0.500,trim=1cm 0cm 0cm 15cm]{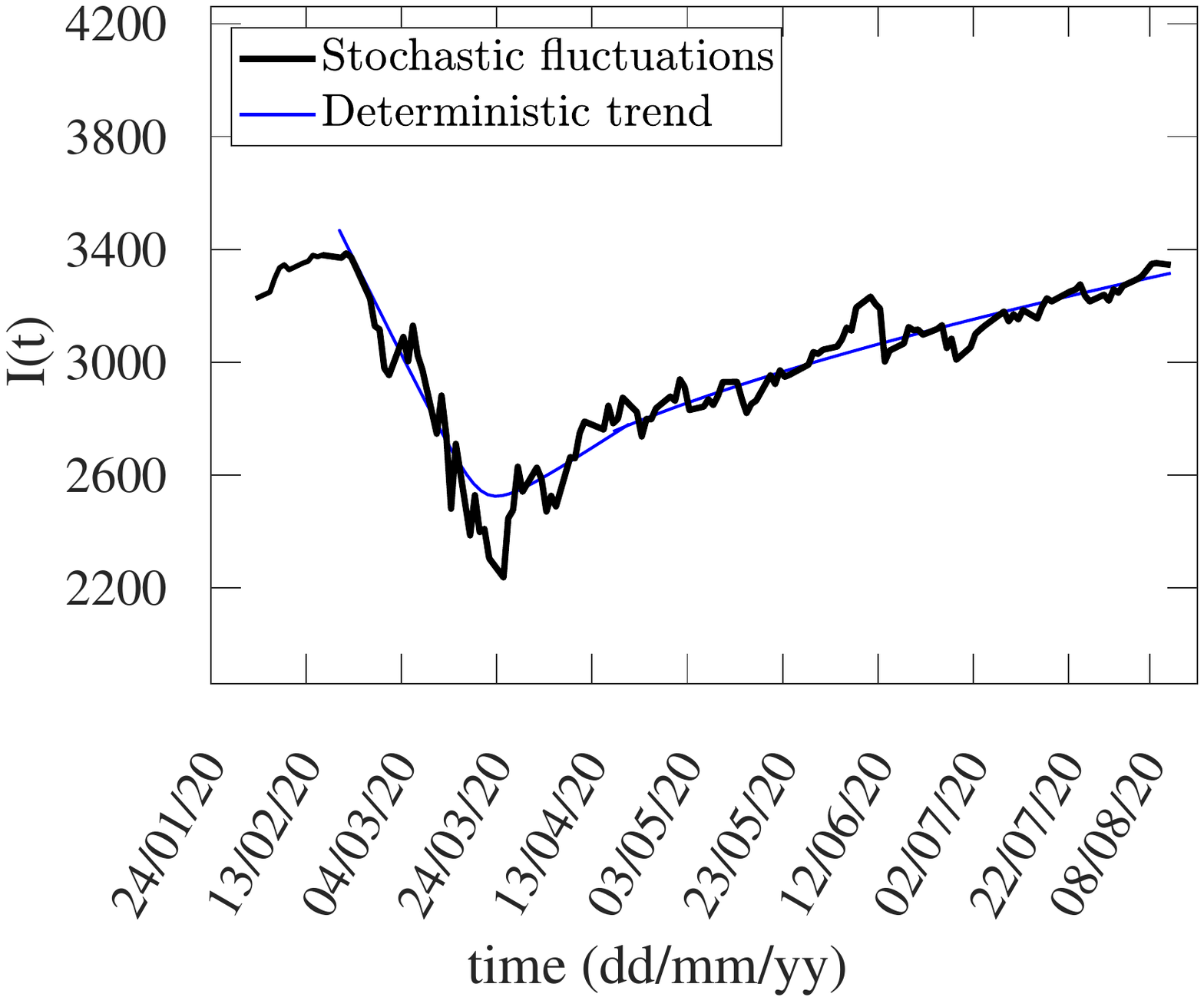}
	\caption{Schematic plot of a downward and upward swing in the S\&P500. The index is decomposed into a deterministic trend (blue) and stochastic fluctuations (dotted black).  }
	\label{fig:Trend_and_fluctuations}       
\end{figure}

The stationary testing methods include an estimation of a Hurst exponent $H$, which itself characterizes the self-similarity of the time series \cite{marton2014detrended}.  Time series are classified as monofractal or multifractal
depending on how the Hurst exponent varies over several orders of magnitude \cite{gu2006detrended}. The monofractal time series posses a unique Hurst exponent. The stationarity test for the monofractal time series is based on its autocorrelation function, where its decay depends directly on the Hurst exponent \cite{kantelhardt2002characterization}. The multifractal time series does not have a unique Hurst exponent \cite{kantelhardt2002multifractal}. The stationarity test for multifractal time series is based on the relation between two sets of scaling exponents obtained from the standard detrending fluctuation analysis (DFA) \cite{maganini2018investigation, kantelhardt2002characterization}  and the generalized detrending fluctuation analysis (G-DFA) \cite{kantelhardt2002characterization,gu2010detrending}, respectively.\par
This paper is divided into four parts: In Section \ref{Non-stationarity of Price return}, the governing equation of the simple price return is presented. Section \ref{Detrended methods} presents a straightforward way to detrend the stock market price, based on the optimization of a moving window via the kurtosis evaluation. In Section \ref{Stationarity of detrended price}, we perform two statistical tests on the detrended price return to show that it is stationary.  The first test is based on the evaluation of the  Hurst exponent in terms of the orders of the statistical moments of the time series.
The second test verifies that the power spectrum of the time series equals the Fourier transform of its autocorrelation \cite{witt1998testing,ivanov2009levels}. Apart from showing that our detrended time series is stationary, we provide the details of the Hurst exponent calculation and the characterization of the detrended price return. Finally, the conclusions and perspectives of this work are presented in Section \ref{Conclusions}.\par

\section{Governing equation of price return} \label{Non-stationarity of Price return}

The conventional approach to option pricing analysis leads to the classical Black-Scholes equation or non-linear Black -Scholes equation (BSE) as the governing equation of price return\cite{barles1998option}. A limitation of the BSE is the assumption that stock prices follow a geometric Brownian motion pattern, ignoring that prices can have sustained trends and even can reach negative values in the real indexes \cite{OIL}. Our governing equation covers BSEs' limitations, allowing the price to fluctuates in a particular direction (trend) and incorporating stationary q-Gaussian noise. The q-Gaussian noise is a better descriptor of the fluctuations in stock markets \cite{alonso2019q}. For this analysis, the price return at time $t$ is defined as \citep{mantegna1995scaling,Johnson2003,Voit2005}:
\begin{equation}
X(t_{0},t)=I(t_{o}+t)-I(t_{o}),
\label{eq:Price return}
\end{equation}
where $I(t_{0})$ is the stock market index at time $t_{0}$, and $I(t)$ is the stock market index for any time  $t > t_{0}$.\par
Our approach decomposes the price return $X(t)$ into a deterministic component $\overline{X}(t)$ and a stationary $q$-Gaussian noise $x$, 
\begin{equation}
X(t)=\overline{X}(t)+x.
\label{eq:detrended}
\end{equation}
In earlier work, we presented the probability density function (PDF) of the detrended price return $x(t)$, described by the functional form \cite{alonso2019q}:
\begin{equation}
P(x,t)=\dfrac{1}{(D t)^{1/ \alpha}}\dfrac{1}{C_{q}} \left[1-(1-q)\dfrac {x^{2}}{(D t)^{2 / \alpha}} \right]^{\dfrac{1}{1-q}},
\label{eq:q-Gaussian}
\end{equation}
where $D$, $q$ and $\alpha$ are time-dependent fitting exponents. The normalization constant $C_{q}$ for $1<q<3$ is given by:
\begin{equation}
Cq =\sqrt{\dfrac{\pi}{q-1}} \dfrac{\Gamma((3-q)/(2(q-1)) ) }{\Gamma(1/(q-1)) }.
\label{eq:Cq}
\end{equation}

Then, we apply the following change of variable $T=t^{\xi}$, where $\xi=\dfrac{3-q}{\alpha}$, in the governing equation of $P(x,t)$ [Eq.(12) from \cite{alonso2019q}]. The corresponding Partial Differential Equation (PDE) is:
\begin{equation} 
\dfrac{\partial P}{\partial T} = D^{\xi} \dfrac{\partial ^{2} P ^{2-q}}{\partial x^{2}}.
\label{eq:PDE}
\end{equation}
By applying Eq.~(\ref{eq:detrended}) the PDF of price return $P(X,T)$ is:
\begin{equation}
P(X,T)=\left( \dfrac{1}{(DT)^{1/\alpha}}\dfrac{1}{C_{q}} \right)\left[1-(1-q)\dfrac {[X-\overline{X}(T)]^{2}}{(DT)^{2/\alpha}} \right]^{\dfrac{1}{1-q}}.
\label{eq:PDF_drift}
\end{equation}

Then, the chain rule is applied to obtain the governing equation of $P(X,T)$.
\begin{equation}
\begin{split}
\dfrac{\partial^{2}}{\partial ^{2} x}&=\dfrac{\partial ^{2}}{\partial ^{2} X} \\
\dfrac{\partial}{\partial T}&= \dfrac{1}{\xi} t^{1-\xi} \dfrac{\partial}{\partial t} + \dfrac{\partial \overline{X}(T)}{\partial T}\dfrac{\partial}{\partial X}\\
\end{split}
\label{eq:dx}
\end{equation}
By replacing Eq(\ref{eq:dx}) in Eq(\ref{eq:PDE}), the general PDE can be defined considering that $\frac{t^{1-\xi}}{\xi}=\frac{ \partial t}{\partial T}$ as:
\begin{equation}
\begin{split}
\dfrac{\partial P}{\partial T}=D^{\xi} \dfrac{\partial^{2} P^{2-q}}{\partial X^{2}}-\dfrac{\partial \overline{X}(T) }{\partial T } \dfrac{\partial P}{\partial X}.  \\
\end{split}
\label{eq:PDE_drift}
\end{equation}

The term $D^{\xi}$  is the diffusive parameter, which is independent on $X$. Consequently, a comparison between Eq(\ref{eq:PDE_drift}) with the linear Fokker-Planck equation (LFPE) is made. The LFPE is :
\begin{equation}
\dfrac{\partial P(X,T)}{\partial \tau}=-\dfrac{\partial (D_{1}P(X,T))}{\partial x}+\dfrac{\partial ^{2}(D_{2}P(X,T))}{\partial x^{2}},
\label{eq:NLFPE}
\end{equation}
After that, the diffusion coefficient in the LFPE is defined as: $D_{2}=D^{\xi}P(X,T)^{1-q}$ . By replacing Eq(\ref{eq:PDF_drift}) into this equation, we obtain an explicit expression for diffusion coefficient of the LFPE: 
\begin{equation}
D_{2}=D^{2/\alpha} C_{q}^{q-1} T^{(q-1)/(3-q)}\left[1-\dfrac{(1-q)[X-\overline{X}(T)]^{2}}{(D^{\xi}T)^{2/(3-q)}} \right]
\end{equation}
and the drift term is:
\begin{equation}
D_{1}=\dfrac{\partial \overline{X}(t) }{\partial T }
\end{equation}
The drift term is the overall direction of the index price (trend), and it shows particular tendencies. It is associated with the systematic risk response, where external events can produce a change of trend direction \cite{karina2020forecasting}. This feature implies that future price changes are dependent on the past values and therefore it is possible to develop a forecast in the short term with some degree of confidence.\par
\begin{figure}[!htbp]
	\centering
	\includegraphics[scale=0.50,trim=2.5cm 0.0cm 1cm 8cm]{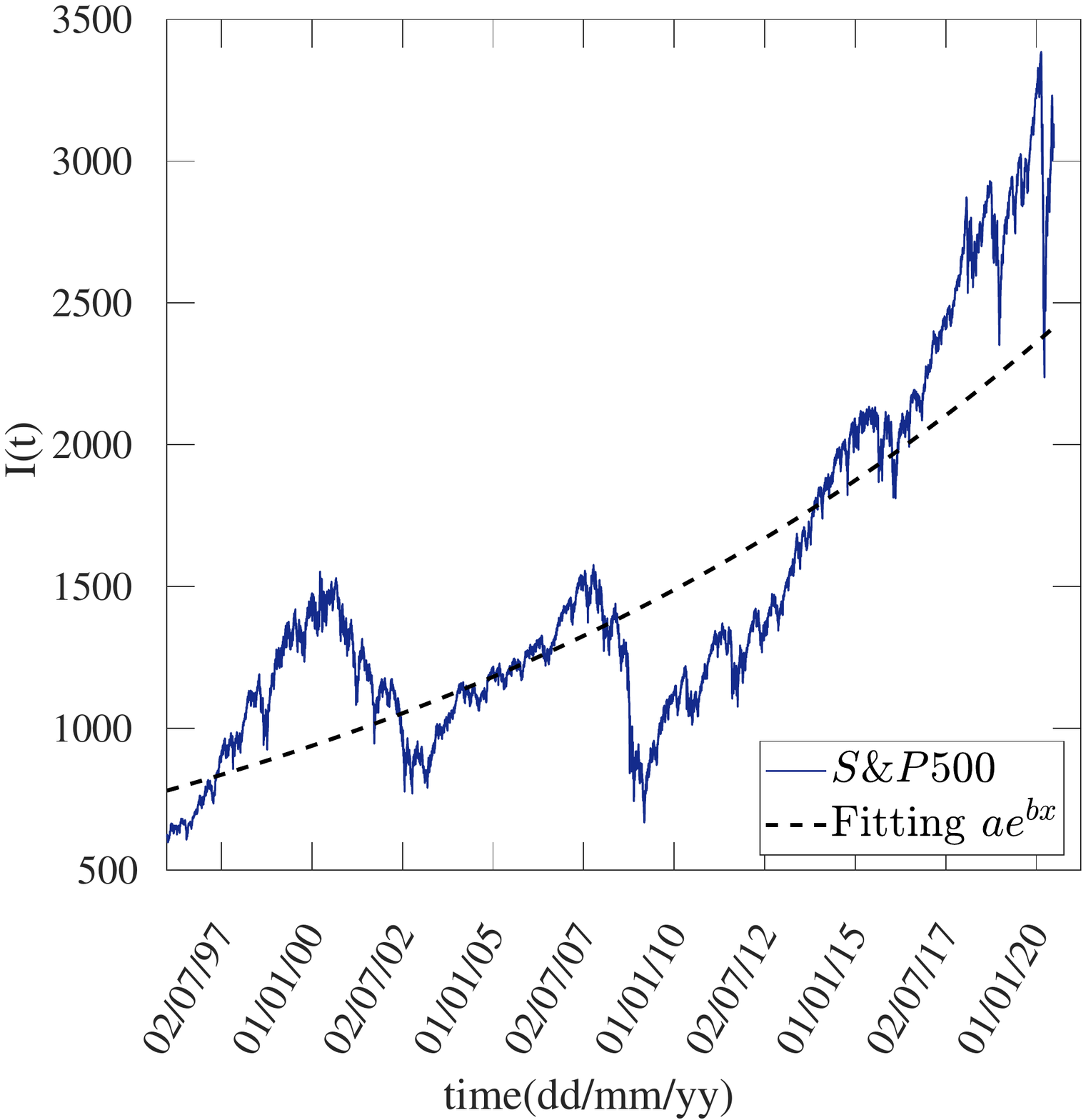}
	\caption{(Stock market index $I(t)$ of S\&P500 in the period of 02/01/96 to 30/06/20 (24 years). Considering the GBM in Eq.~(\ref{eq:BM}), the deterministic part is modeled to find the trend of the S\&P500. As a result, the  fitted line $ae^{bx}$ is an approximation of the  S\&P500 data trend, where $a=7.45(10^{-38})\pm5.97(10^{-39})$ and $b=0.0001264\pm10^{-07}$. However, if this approximated trend is subtracted, the fluctuations will form a  not stationary time series because their statistical properties will change on time. }
	\label{fig:Time_series}      
\end{figure}

The Fokker-Planck equation in Eq.~(\ref{eq:NLFPE}) is used to obtain the probability distribution of price return at any time. The stochastic differential equation (SDE) represents the dynamics of price return as a function of time {\cite{vasconcelos2004guided}}. The SDE can be obtained by applying Ito Lemma \cite{vasconcelos2004guided} in Eq.(\ref{eq:NLFPE}):

\begin{equation}
dX_{T}=\mu(T) dT + \sigma(X,T) dW_{T},
\label{eq:KBM}
\end{equation}
with drift $\mu(T)=D_{1}(T)$ and variance  $\sigma^{2}(X,T)=2D_{2}(X,T)$. The function $W_{T}$ is the q-Gaussian noise. The interpretation of our SDE is that in a small time interval, $dT$, the price return changes its value due two effects: the first one is related with the drift $D_{1}$ that depends on time only. The second is the diffusion term $D_{2}$ that depends on time and price return. The SDE in Eq.~(\ref{eq:KBM}) is different from the Geometrical Brownian Motion (GBM) \cite{TsayRueyS.2005AoFT}:
\begin{equation}
dI_{T}=\mu I_{T}dT + \sigma \,I_{T} dB_{T},
\label{eq:BM}
\end{equation}
where, $\mu$ and $\sigma$ values are constants, and $B_{T}$ is the standard Brownian noise. The GBM is the basic model for stock price dynamics in the Black-Scholes framework\cite{vasconcelos2004guided}. If we consider  the approximation, $Z_{T}=dI_{T}/I_{T}$, where $Z_{T}$ is a logarithmic difference of the stock market index $Z(t_{0},t) =\ln \left[ \frac{I(t_{0}+t)} {I(t_{0})}\right] $. Then, Eq.~(\ref{eq:BM}) can be simplified as $Z_{T}=\mu dT + \sigma \, dB_{T}$.  The solution of this equation is an exponential growth superposed by a normally distributed Brownian process \cite{vasconcelos2004guided}. This solution constitutes a rough approximation of the price return, as shown in Figure \ref{fig:Time_series}. In some studies, the log-return $Z_{T}$ is often used for technical analysis because it is close to the percentage price change  \citep{ren2018different,CaoWen2017}, a concept often used to neutralize most of the non-stationary effects \citep{gopikrishnan1999scaling,Mandelbrot1997,Cont2004,Jondeau2007}. \par
 One of the key properties of the GBM approach to modeling asset prices is that prices will not go negative, an assumption that may not be reasonable given recent events in some asset markets. Some examples are crude oil's price fall and the bankruptcy of some assets of global energy companies, which had reached negative values \cite{OIL}. Additionally, in the GBM, it is assumed that the natural logarithmic
of price return follows the Brownian motion with drift and that the price return has a normal distribution \cite{vasconcelos2004guided}. However, price returns or log price returns have been modeled by a q-Gaussian distribution function \cite{xu2016transition}. Thus, the stochastic differential equation, Eq.(\ref{eq:KBM}), could potentially model this behavior better than the standard GBM.\par
In what follows, we present an analysis of the S\&P500 index per minute for the past $24$ years in the time frame of 1996-2020 as a case study to decompose the time series into a  deterministic trend and a stationary stochastic part. The S\&P500 stock market index data was obtained from the Thomson Reuters Corporation \cite{Reuters2020}. Before the analysis, few artifacts were removed from the data \cite{Zen2020,CNNfn2000,ferreira2015earthquakes}, as explained in the Appendix \ref{AppendixA:Cleansing-data}.

\section{Detrending time series} \label{Detrended methods}
The Moving Average is a widely known method to remove the trend in a time series \citep{6708545,gu2010detrending,alonso2019q}. Many variations and implementations have been developed by researchers, but its underlying purpose remains, which is to determine the trend. Despite new models, the MA method is still considered as a good option for detrending the data due to its easiness, objectiveness, reliability, and usefulness \citep{6708545}.

The MA is a common average within a time window size that is shifted forward until the end of the data set  \citep{6708545}. Each point in the time series data is equally weighted. The arithmetic centered moving average has been the most widely used MA case \citep{marathe2005validity}. 

The MA method offers different options for its calculation. The first option is regarding how the time window will be shifted until the end of the data. For this analysis, it has been used an overlapped or rolling window. These windows are extended over an interval $[t, \, \, \, t + t_{w}]$, where the consecutive window is
incremented by one time point to: $[t+1, \,\,\, t+t_{w}+1]$. The $t_{w}$ represents the size of the time window.
The second option is choosing the polynomial order $k$ for fitting each time segment for the detrending process. The original MA uses $k=0$, which represents an unweighted mean \citep{peng1994mosaic}. The option $k=1$ is used in other studies when the focus is on the local trend (particular tendency per window) \citep{alessio2002second}. These options were tested and applied in the index price $I(t)$. In our analysis, there is no major difference between  $k=0$ and $k=1$, so that $k=0$ was used.

\subsection{Optimal time window} \label{Optimaltw}
The size of the time window,  $t_{w}$, is the key parameter to obtain stationary detrended data. The distribution of the data set of each window $t_{w}$ should be similar, without outliers (extreme events) to achieve an equilibrium state of the dynamic system \cite{xu2016transition}. Xu et al.\citep{xu2016transition} applied the kurtosis definition to compare each distribution of the data set of each window $t_{w}$.
The kurtosis is defined as the fourth standardized moment, defined as the relation between central moments $\langle u^{n} \rangle$ of order $n$,

\begin{equation}\label{eq:Kurtosis}
K=\frac{\langle u^{4} \rangle}{\langle u^{2} \rangle^2},
\end{equation}
where the delimiters $\langle \, \rangle $ denotes time average for each particular window.\par 
The optimal time window must satisfy the condition of $K\approx3$, to obtain a distribution, in which extreme or outlier events are very unlikely to occur \citep{buonocore2019interplay,xu2016transition}.
\begin{figure}[!htbp]
	\centering
	\includegraphics[scale=0.45,trim=1cm 0.0cm 1cm 15cm]{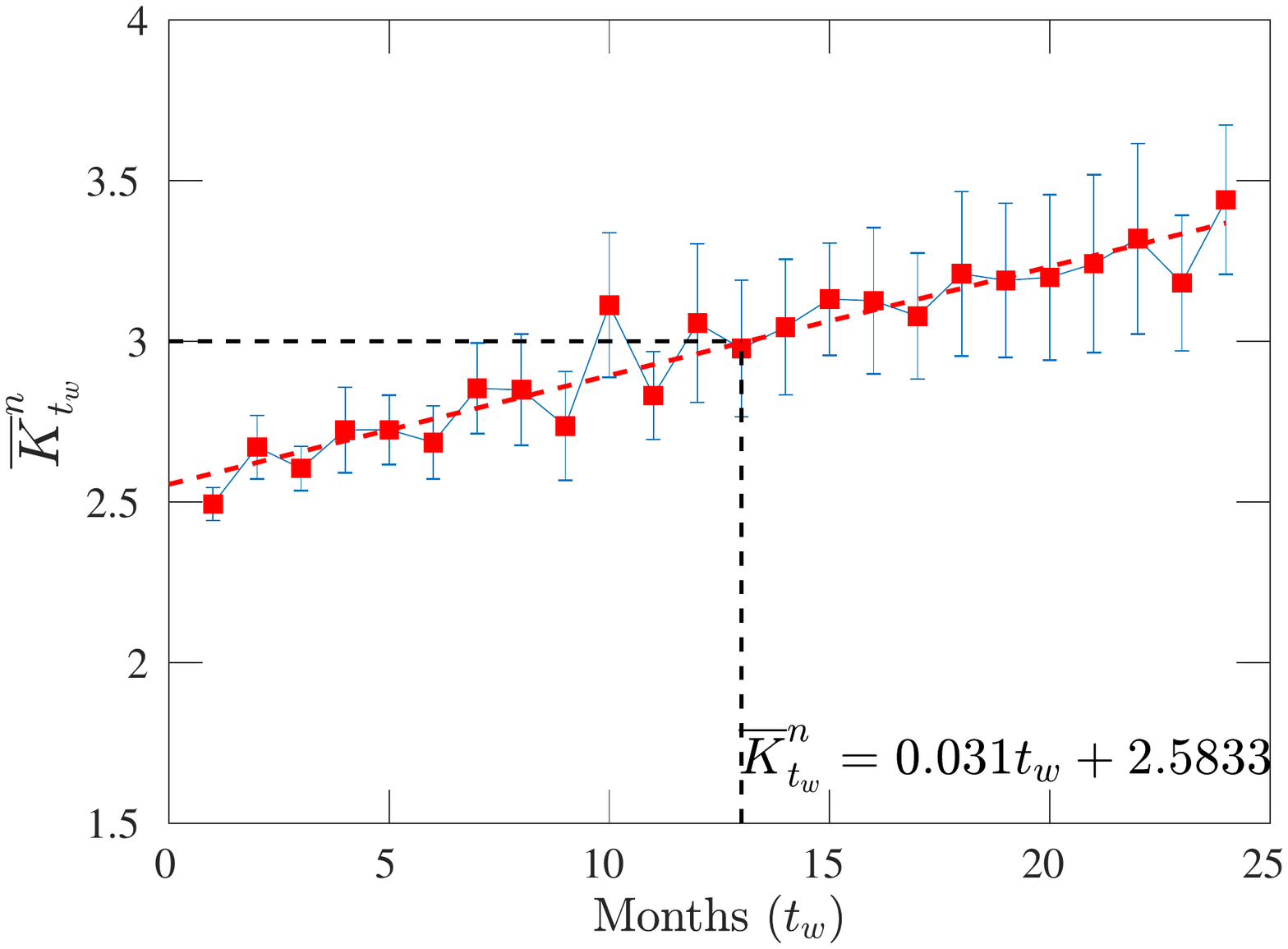}
	\caption{Optimal window size $t_{w}$  that will be used for the Moving Average. The  $\overline{K}_{tw}^{\,n}=3$ yields a time window of $13\pm 1$ months. In order to have an integer unit for $t_{w}$ , $t_{w}$ = 12 months was chosen. The blue bars are graphical representations of the standard error of $\overline{K}_{tw}^{\,n}$ data in average per $t_{w}$}
	\label{fig:Kurtosis}      
\end{figure}
\\
Before applying the MA method, we developed a preliminary process to choose an optimal time window $t_{w}$. We used a range of values between $1$ to $25$ months as a possible time window. For each possible $t_{w}$ value, the total length of $N$ of the time series is split into non-overlapping time windows. The total length of $N$ is often not a multiple of the time window, $t_{w}$. Consequently,  the number of segments is defined by rounding them to the nearest lower integer $ \lfloor n=N/{t_{w}} \rfloor $. Next, the kurtosis in each $j^{th}$ window is calculated, where $j= 1,2,3,4.....n$. and averaged for the $n$ windows as follows;
\begin{equation}\label{eq:Kurtosis2}
\overline{K}_{t_{w}}^{\,n}=\frac{1}{n}{\sum_{j=1}^{n} K_{t_{w} }(j)},
\end{equation}
where  $\overline{K}_{t_{w}}^{\,n}$  denotes the average over the $n$ windows. Figure \ref{fig:Kurtosis} presents the results of $\overline{K}_{t_{w}}^{\,n}$. The optimal window size is $13\pm 1$ months. The time window assumed for the MA analysis is $12$ months. The smaller value of time window, $t_{w}$, is chosen because it will preserve features of the data such as peak height and
width, which is usually attenuated with a longer time window $t_{w}$.

\subsection{Calculation of trend and fluctuating part}
\label{sec:MA}
The non-stationary time series of the stock market index $I(t)$ is splited into a trend  $\tilde{I}(t)$ and a stationary fluctuating component $I^{*}(t)$,
\begin{equation}\label{eq:MA4}
I^{*}(t)=I(t)-\tilde{I}(t).
\end{equation}
The trend $\tilde{I}(t)$ is constructed for the index price of S\&P500 applying MA method with $k=0$ and considering overlapped windows of $t_{w}=12$ months. This process is constructed based on a weighted sum or average of each time window. There are three cases to calculate the moving average that depends on the position of the moving window \citep{gu2010detrending}:

\begin{description}[font=$\bullet$~\normalfont]
	\item For $ t< \frac{t_{w}}{2}$
	\begin{equation}\label{eq:MA1}
	\tilde{I}(t)=\frac{1}{t_{w}}{\sum_{k=-\lfloor(t-1)\rfloor}^{\lceil(t_{w}-1)/{2}\rceil} I(t+k)}
	\end{equation}
	\item For $\frac{t_{w}}{2}<t<N-\frac{t_{w}}{2}$
	\begin{equation}\label{eq:Ma2}
	\tilde{I}(t)=\frac{1}{t_{w}}{\sum_{k=-\lfloor(t_{w}-1)/{2}\rfloor}^{\lceil(t_{w}-1)/{2}\rceil} I(t+k)}
	\end{equation}
	\item For $t>N-\frac{t_{w}}{2}$ 
	\begin{equation}\label{eq:MA3}
	\tilde{I}(t)=\frac{1}{t_{w}}{\sum_{k=-\lfloor(t_{w}-1)/{2}\rfloor}^{\lceil N-t\rceil} I(t+k)},
	\end{equation}
\end{description}
with the time step of $t=1,2,3....N$ for the index fluctuations.

\begin{figure}[!htbp]
	\centering
	\includegraphics[scale=0.5,trim=2.0cm 6.0cm 2cm 9cm]{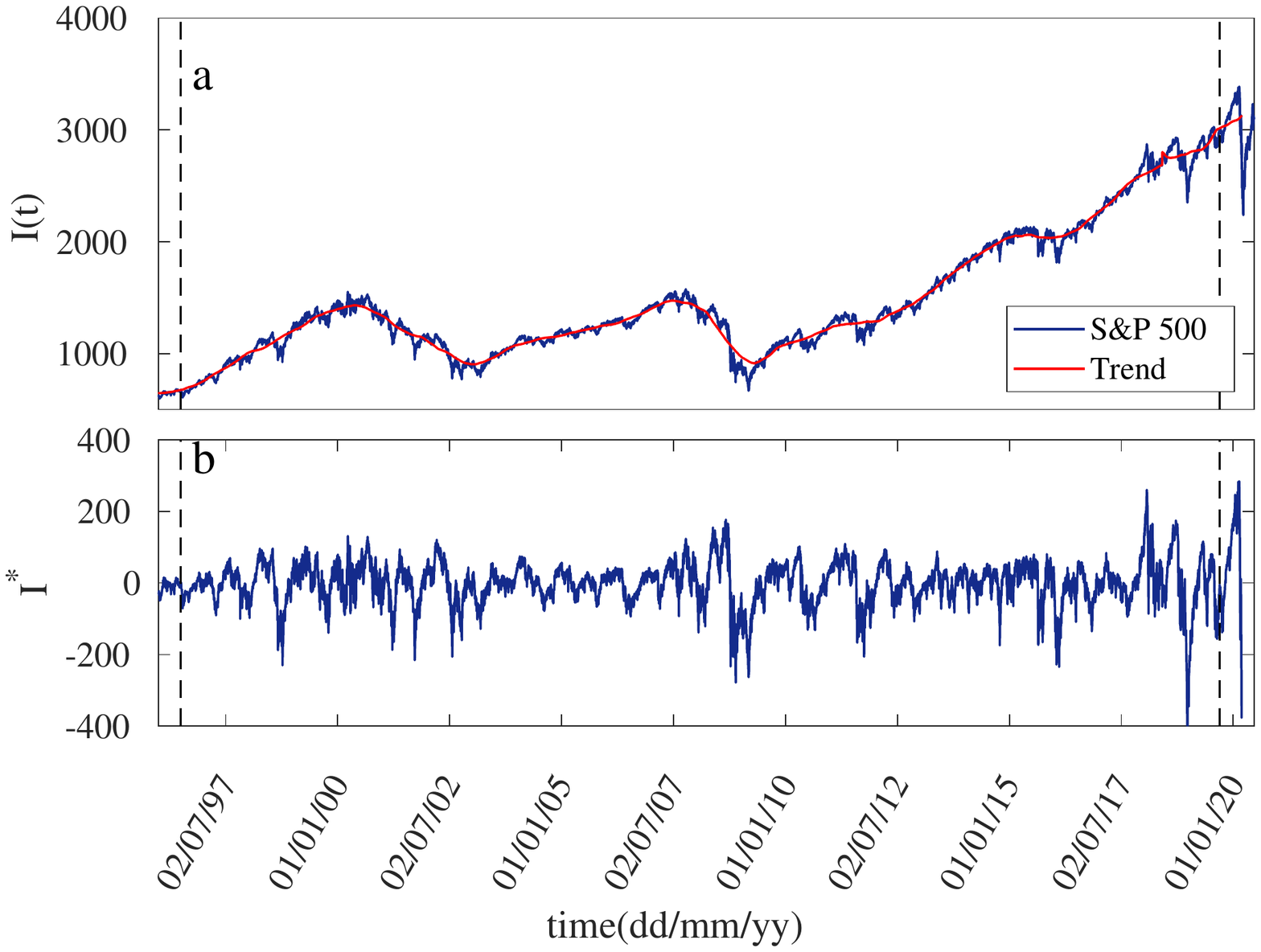}
	\caption{ (a) Trend obtained after applied the MA analysis for a time window of 12 months for S\&P 500 data. (b) Detrend price $I^{*}(t)$ after subtracting the trend shown on subfigure 4-a.}
	\label{fig:Price_detrended}       
\end{figure}

Figure \ref{fig:Price_detrended}a presents the trend $\tilde{I}(t)$ after applying the MA analysis for a time window of 12 months. The trend obtained represents a general tendency of the stock market index $I(t)$.  The Figure \ref{fig:Price_detrended}b shows the detrended price as a result of subtracting the obtained trend.

\section{Stationarity of detrended price return}  \label{Stationarity of detrended price}
In the following section, we test the detrended time series for stationarity. These tests are based on the calculation of the Hurst exponent and the power spectrum analysis, respectively.
The detrended price return is defined as,
\begin{equation}
x(t)=I^{*}(t_{o}+t)-I^{*}(t_{o}),
\label{eq:Index}
\end{equation}
where $I^{*}(t_o)$ is the detrended stock market index at time $t_o$, and $I^{*}(t)$ is the detrended stock market index for any time  $t>t_o$.\par
\subsection{First stationary test}
For this test, the Hurst exponent of this time series is calculated. This Hurst exponent is a measure of long-range memory of a time series. More specifically, it measures the statistical dependence of two points of the time series with respect to the time difference between them \cite{kantelhardt2002multifractal}. The Hurst exponent is better known as the \textit{index of dependence}, and as such, it is a dimensionless estimator of self-similarity in a time series \cite{marton2014detrended,horvatic2011detrended}. The self-similarity occurs when a structure repeats itself on subintervals, so they are \textit{scale-invariant}, which is a property that some time series posses \cite{marton2014detrended}. Two types of self-similar signals exist. The monofractal signal, which presents a unique scaling behavior of complexity.  The scaling behavior is obtained from the ratio of the self-similar pattern that changes with the scale at which it is measured. The second type is the multifractal signal, which has more than one scaling behavior \citep{hu2001effect}.\par
The Hurst exponent is estimated by applying the detrended fluctuation analysis (DFA) \citep{hu2001effect} explained in the Appendix \ref{AppendixB:DFA}. The DFA is a technique to demonstrate self-similarity in a time series \cite{ausloos2012generalized}. The DFA method provides complementary information, such as the Hurst exponent. Additionally, by analyzing the statistics, it is possible to characterize the time series as monofractal or multifractal if the self-similarity exists.\par
The first part of the DFA analysis consists in removing the trend of the original time series considering non-overlapping segments. However, our trend was removed by applying a simple but still rigorous MA method considering overlapped time windows with an optimal size of time window $t_{w}$.\par
The DFA is based on the calculations of the \textit{statistical moments}, $F_{\mathrm{w}}(s)$, as a function of the  \textit{time segment} $s$, where $\mathrm{w}$ is the order of the moment \citep{peng1994mosaic,liu1999statistical,alessio2002second}. The full definition is given in Eq.(\ref{eq:DFA3}) in the Appendix \ref{AppendixB:DFA}. If the time series satisfies the $F_{\mathrm{w}}(s)\sim s^H$, the time series is monofractal, and the exponent $H$ is called the Hurst exponent \citep{kantelhardt2002multifractal,kantelhardt2001detecting}. Other time series do not have a unique, constant valued $H$.  Their scaling exponent $H$ depends on the order of the moment $\mathrm{w}$ (Figure \ref{fig:Moments}). These are classified as a  multifractal time series. The generalization method of DFA  (G-DFA) is tailored for multifractal time series. In the G-DFA, the \textit{generalized statistical moments} are defined by $G_{\mathrm{w}}(s)$ in Eq.~(\ref{eq:Ftau1}). If the series is multifractal, the exponent $\tau({\mathrm{w}})$ is the so-called \textit{generalized scaling exponent} (GSE). The multifractal time series satisfies the scaling law of $G_{\mathrm{w}}(s) \sim s^{\tau(\mathrm{w})}$. For stationary and normalized multifractal time series the generalized scaling exponent $\tau(\mathrm{w})$ can be defined in terms of the generalized Hurst exponent $h(\mathrm{w})$ \citep{kantelhardt2002characterization}:
\begin{equation}
\tau(\mathrm{w})=\mathrm{w}h(\mathrm{w})-1.
\label{eq:Tau}
\end{equation}

\begin{figure}[!htbp]
	\centering
	\includegraphics[scale=0.43,trim=0cm 0.3cm 0cm 15cm]{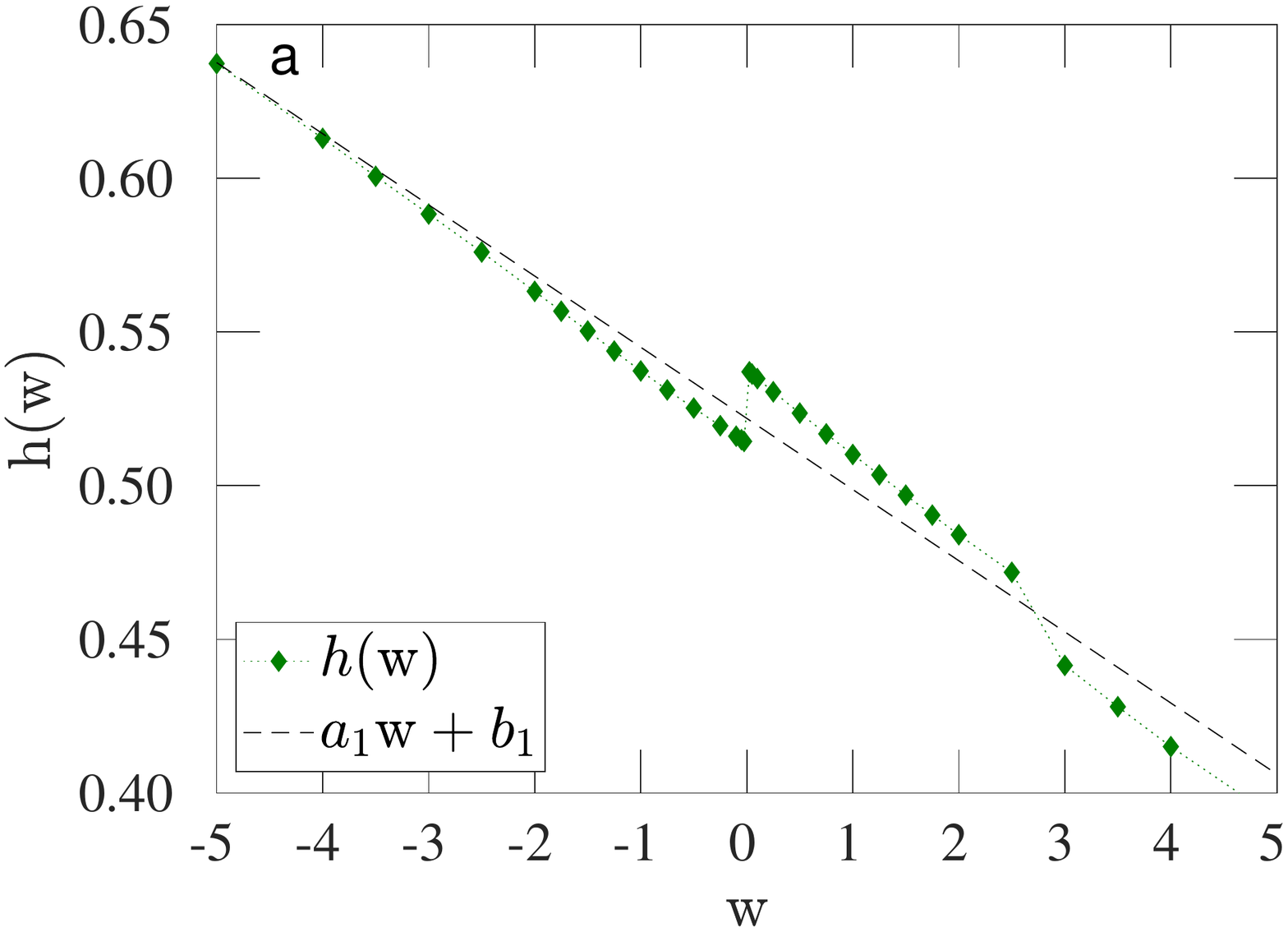}
	\includegraphics[scale=0.43,trim=0cm 0.3cm 0cm 15cm]{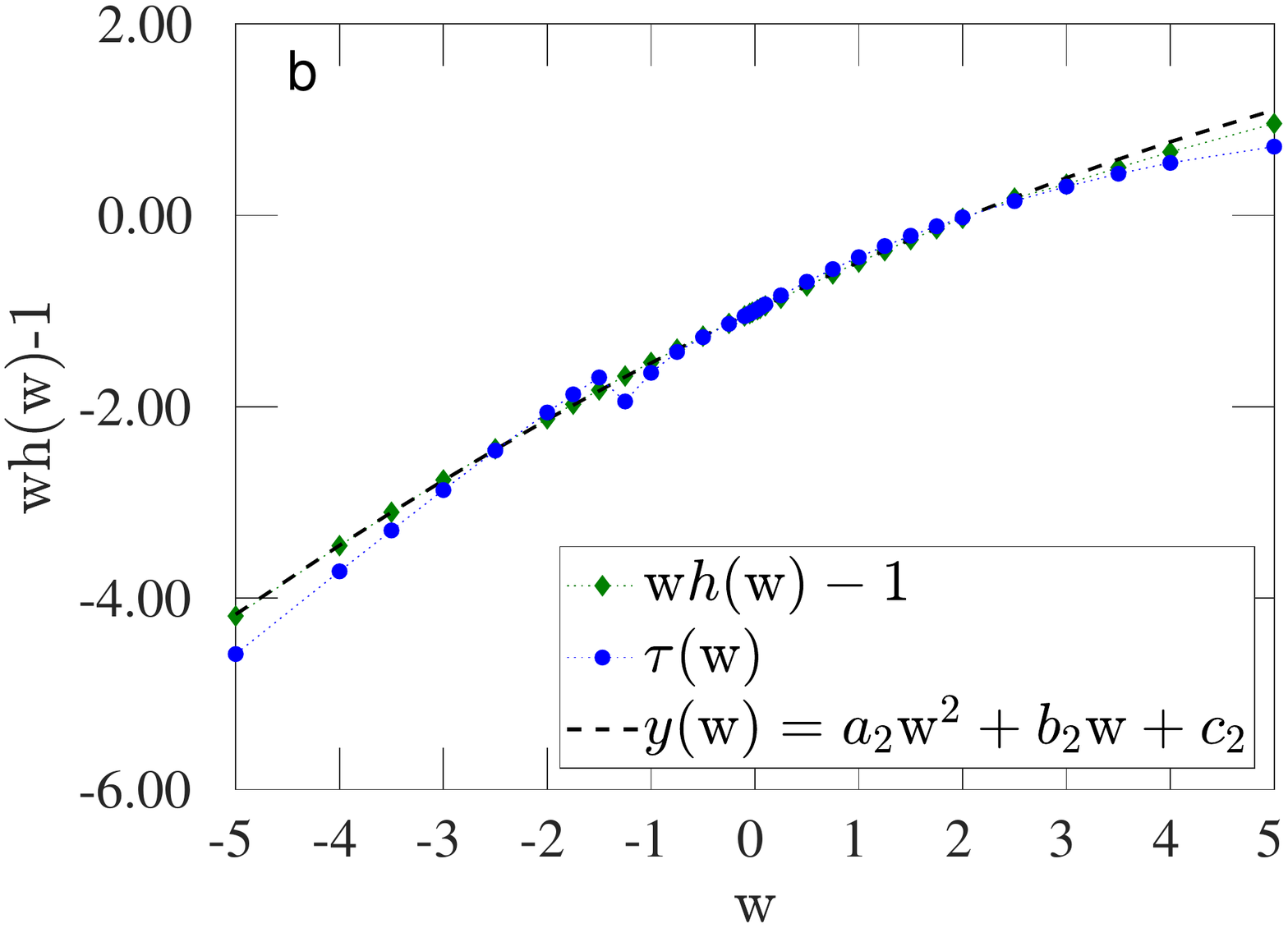}
	\caption{Evaluation of the scale exponents, $h({\mathrm{w}})$, and the generalized scaling exponent, $\tau({\mathrm{w}})$, obtained from results shown in Figure \ref{fig:Moments} and \ref{fig:MomentsZq} respectively. The generalized Hurst exponent $h(\mathrm{w})$ follows Eq.~(\ref{eq:Tau}) for a  normalized and multifractal time series.(a) For the fitting of the generalized Hurst exponent  $a_{1}=-0.0232\pm0.002$ and $b_{1}=0.5219\pm0.004$ (b) For the fitting of the generalized scaling exponent $a_{2}=-0.02157\pm  3.8(10^{-4})$, $b_{2}= 0.527 \pm 0.0012$, $c_{2}=-0.9954\pm0.0034$ }
	\label{fig:tau_q}       
\end{figure}

The Figure \ref{fig:tau_q}a shows the generalized Hurst exponent obtained from Figure \ref{fig:Moments}, and its fitting. The generalized Hurst exponent is approximated to $h(\mathrm{w})  \simeq -0.023\mathrm{w}+ 0.521$ for $-5.0 \leq \mathrm{w} \leq5.0$. Figure \ref{fig:tau_q}b shows the $\tau(\mathrm{w})=\mathrm{w}h(\mathrm{w})-1$ relationship, proving that the detrended price return is stationary and multifractal  \citep{kantelhardt2002characterization}. \par

\subsection{Second stationary test}

This test is presented to prove stationary without using the Hurst exponent. This test is based on the autocorrelation and the power spectrum analysis of the time series, by applying the stationary definition  \citep{witt1998testing}.\par
\begin{figure}[!htbp]
	\centering
	\includegraphics[scale=0.43,trim=0cm 0.0cm 0cm 14.5cm]{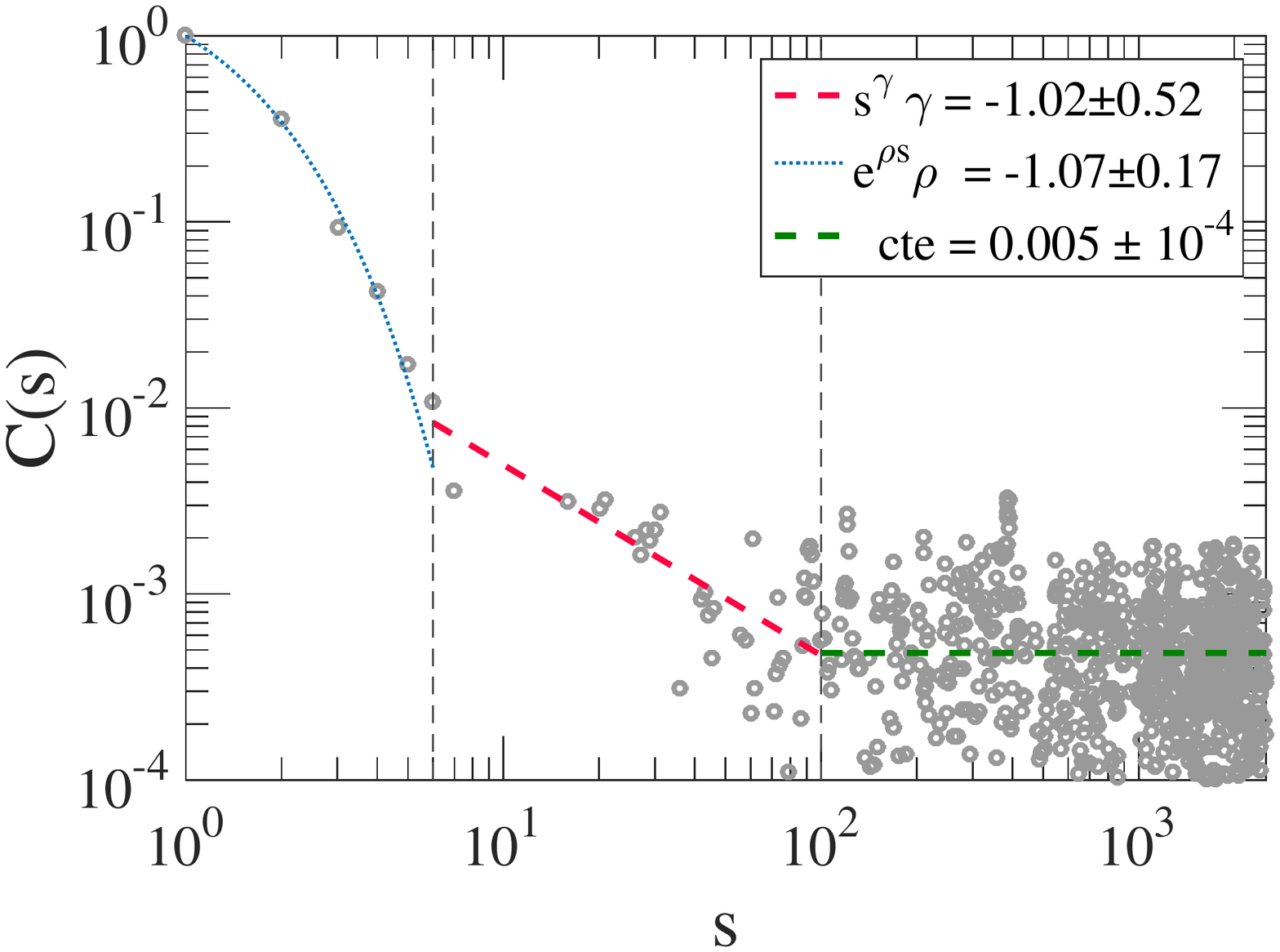}
	\caption{Autocorrelation of the detrended price return calculated as $C_{x}(s)\equiv \dfrac{\langle x_{t} x_{t+s} \rangle}{\sigma_{x}^{2}}$. Short-time correlations are observed for the first $6$ minutes. A weak long-time correlation appears after $100$ minutes.}
	\label{fig:PDF_evolution}       
\end{figure}

First, we proceed to evaluate the correlation of the detrended price return \citep{kantelhardt2002multifractal,gao2003principal,kantelhardt2002characterization,rangarajan2000integrated}. The autocorrelation is defined as;
\begin{equation}
C_{x}(s)\equiv \dfrac{\langle x_{t} x_{t+s} \rangle}{\sigma_{x}^{2}},
\end{equation}
\begin{equation}\nonumber
{\sigma_{x}^{2}}=\langle (x_{t}-\mu)(x_{t}-\mu) \rangle.
\end{equation}
The presence of short-time autocorrelations is shown in  Figure \ref{fig:PDF_evolution}, where the detrended price return for a specific time is represented by $x_{t}$. The power-law fitting of the autocorrelation holds only one order of magnitude. For lower values of $s$, the autocorrelation has an exponential trend, and higher values show a weak but non-zero correlation. A transition zone is observed between lower and higher time-lag values, and it decreases rapidly with a power law that has an exponent $\gamma=-1.02\pm0.52$. 

\begin{figure}[!htbp]
	\centering
	\includegraphics[scale=0.45,trim=1cm 0.0cm 0cm 14.5cm]{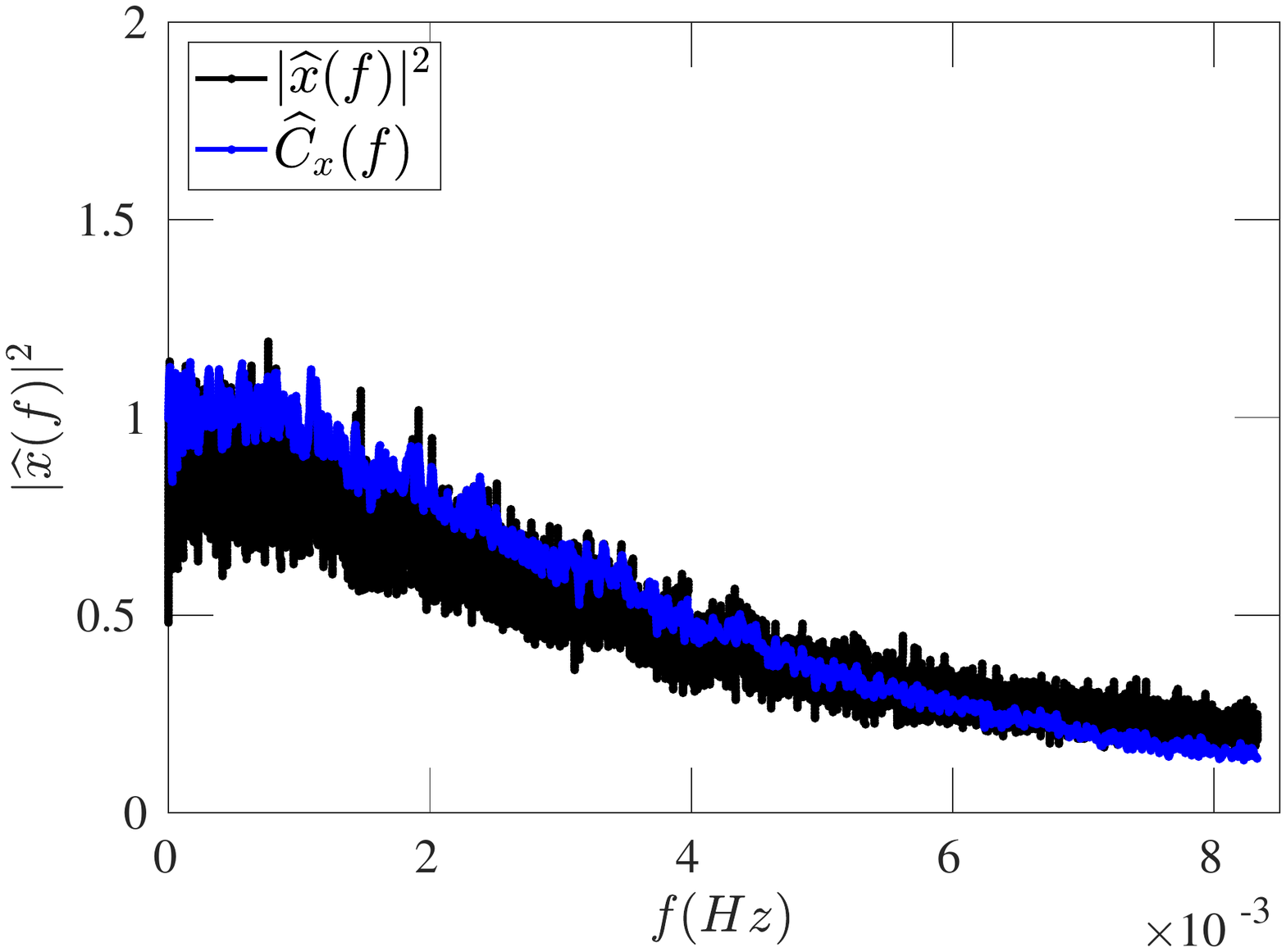}
	\caption{The Power-spectral density $|\widehat{x}(f)|^{2}$ of the detrended price return matches with the Fourier transform of its autocorrelation. }
	\label{fig:PS_Price return}       
\end{figure}
\begin{figure}[!htbp]
	\centering
	\includegraphics[scale=0.45,trim=1cm 0.0cm 0cm 14.5cm]{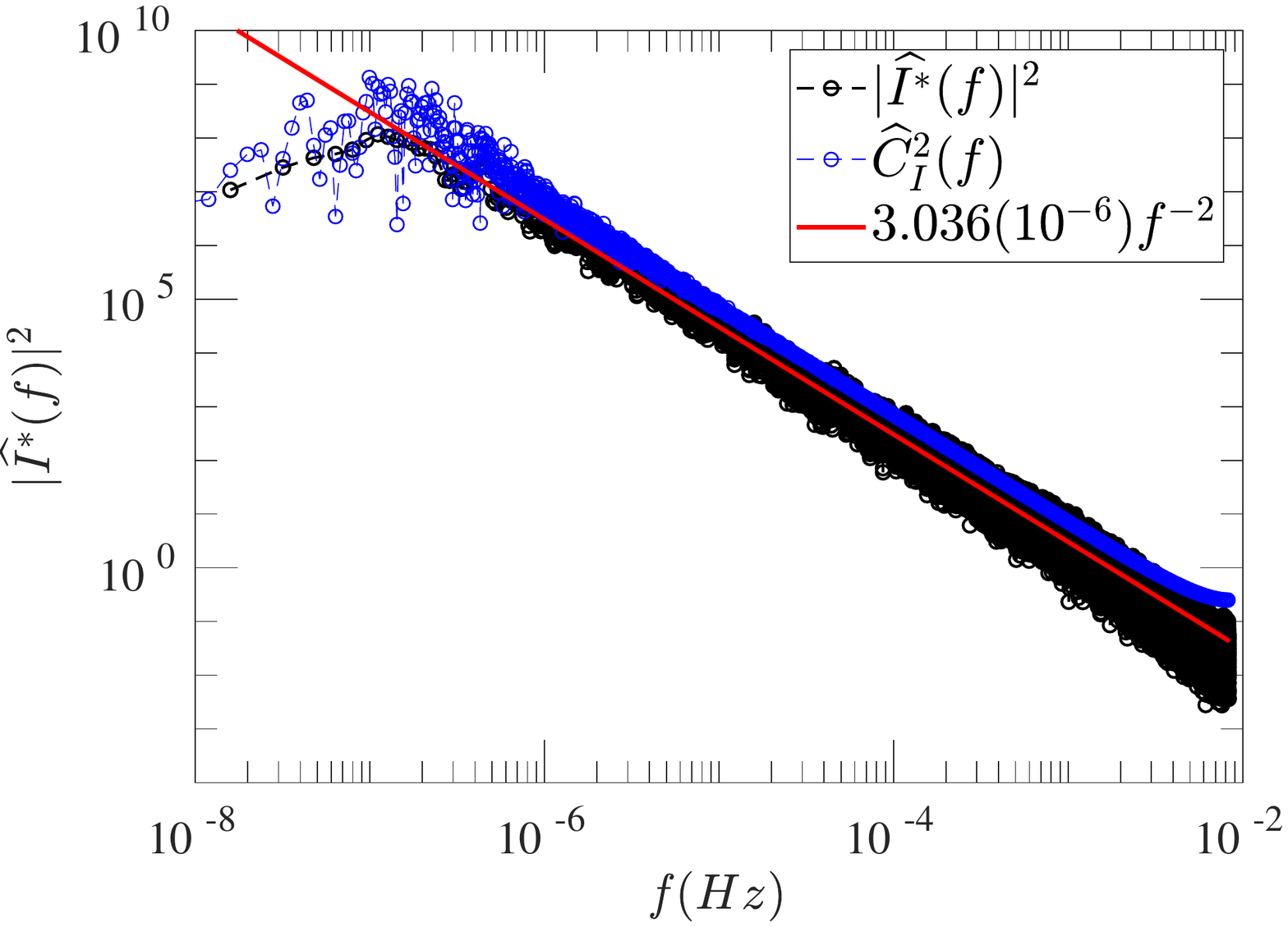}
	\caption{The power-spectral density of the detrended index $I^*(t)$ holds a power law $P(f)= {M}/{f^{2}}$, where $M=3.036(10^{-6})\pm3(10^{-9})$. The following relation $|\widehat{I}(f)|^{2} =\widehat{C}_{I}^{2}(f)$ is observed, suggesting that the Index once detrended is stationarity as well. }
	\label{fig:SP500_Test}       
\end{figure}


Then, we represent $\widehat{x}(f)$ as the Fourier Transformation (FT) of $x_{t}$.
\begin{equation}
\widehat{x}(f)=\dfrac{1}{\sqrt{T}}\sum_{t=0}^{T} x(t) e^{-2 \pi i f t} .
\end{equation}
This test states that any process is stationary if the power spectrum $|\widehat{x}(f)|^{2}$ exists, and can be expressed as the Fourier transform of the autocorrelation ${C(s)}$ \cite{witt1998testing,krenk1983stationary},
\begin{equation}
|\widehat{x}(f)|^{2} =\widehat{C}_{x}(f),
\label{eq:FT}
\end{equation}
We determine that the detrended price return is a stationarity time series based on the results shown in Figure \ref{fig:PS_Price return}, where the match between $|\widehat{x}(f)|^{2}$ and $\widehat{C}_{x}(f)$ is achieved. The $|\widehat{x}(f)|^{2}$ and the $\widehat{C}_{x}(f)$ have been smoothed by applying the Savitzky-Golay filtering. The smoothed is made with the purpose of reducing the noise of the data without distorting their tendency \cite{guinon2007moving,gorry1990general}. The smoothing obtained by the Saviztky-Golay filter is based on the least-squares of a linear fitting across a moving window by applying the normalized convolution integers of Savitzky-Golay, $C_{i}$ \cite{guinon2007moving,gorry1990general}. For this case the length of the moving window is $3.97(10^{-5})$ $Hz.$
Also, the relation $|\widehat{I}(f)|^{2} =\widehat{C}_{I}^{2}(f)$ is observed in Figure \ref{fig:SP500_Test}, suggesting the profile of the detrended price return is stationary too.
From this test, we can conclude once again that the detrended price return is a stationary time series.\par

\section{Conclusions} \label{Conclusions}
To conclude, the S\&P500 index was detrended using a weighted average method with an overlapped time window of one year. Two independent tests demonstrated that the detrended price return is a multifractal, stationary time series with short-time correlations.\par
The two independent tests that prove stationarity of the detrended price return were: (1) The generalized scaling exponent $\tau(\mathrm{w})$ can be expressed in terms of the general form of the Hurst exponent $h(\mathrm{w})$, and (2) the power spectrum of the time series is approximately equal to the Fourier transform of its autocorrelation. These tests validated our improved version of the Black-Scholes equations postulated at the beginning of our paper.\par
The governing equations proposed in this paper remove the limitation of the classical Black-Scholes equation and its modifications. By proposing a  simpler and more effective approach, the new governing equations can describe the price return and its characteristics features. The proposed SDE in Eq.~(\ref{eq:KBM}) presents its deterministic component as endogenous or exogenous predictive factors, which is associated with internal/external events that produce changes of trend direction. The stochastic part is represented by the q-Gaussian noise. Our stochastic differential equation captures the index behavior more precisely by having a trend that fluctuates deterministically. Then, the detrended part is presented as a stationarity time series.  Future work should be focused on the analysis of the fractional Fokker-Planck Equation (FFPE) to model anomalous fluctuations of the stock market indexes. A wide range of q-Gaussian and the Levy-Stable distribution functions can be obtained to model the stochastic fluctuations after applying the different definitions of the fractional derivatives available in the market.

\section{Acknowledgement}
We acknowledge the Australian Research Council grant DP170102927. K.A.C. thanks The Sydney Informatics Hub at The University of Sydney for providing access to HPC-Artemis for financial data processing. We thanks Sornette, Tsallis and Christian Beck for inspiring discussions.

\bibliographystyle{cas-model2-names}
\bibliography{GoverningEquation}

\appendix
\counterwithin{figure}{section}
\section{Appendix A: Cleansing Data}
\label{AppendixA:Cleansing-data}
\begin{figure}[!htbp]
	\centering
	\includegraphics[scale=0.28,trim=0cm 5cm 0.0cm 4cm,clip=true]{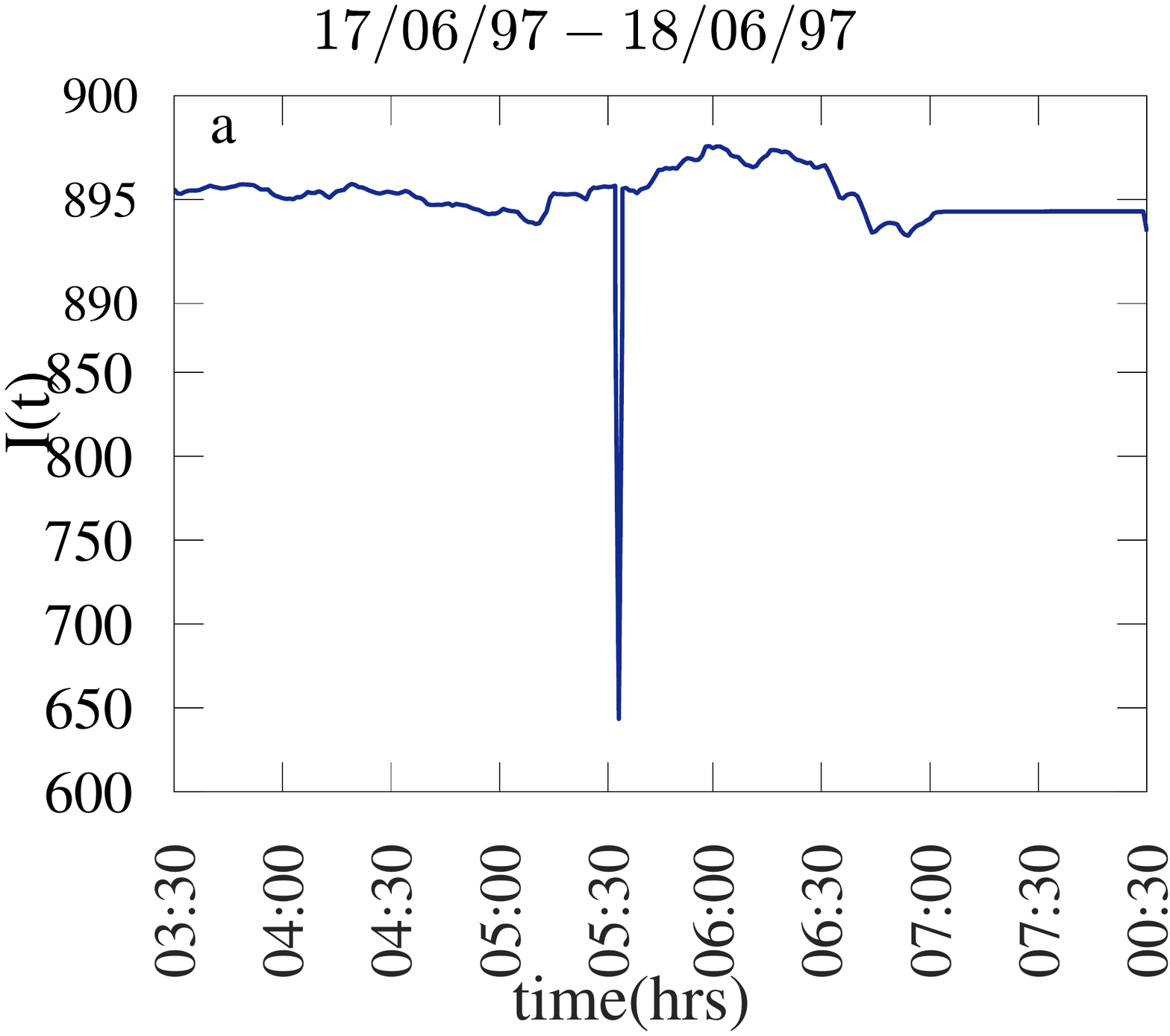}
	\includegraphics[scale=0.28,trim=0cm 5cm 0.0cm 4cm,clip=true]{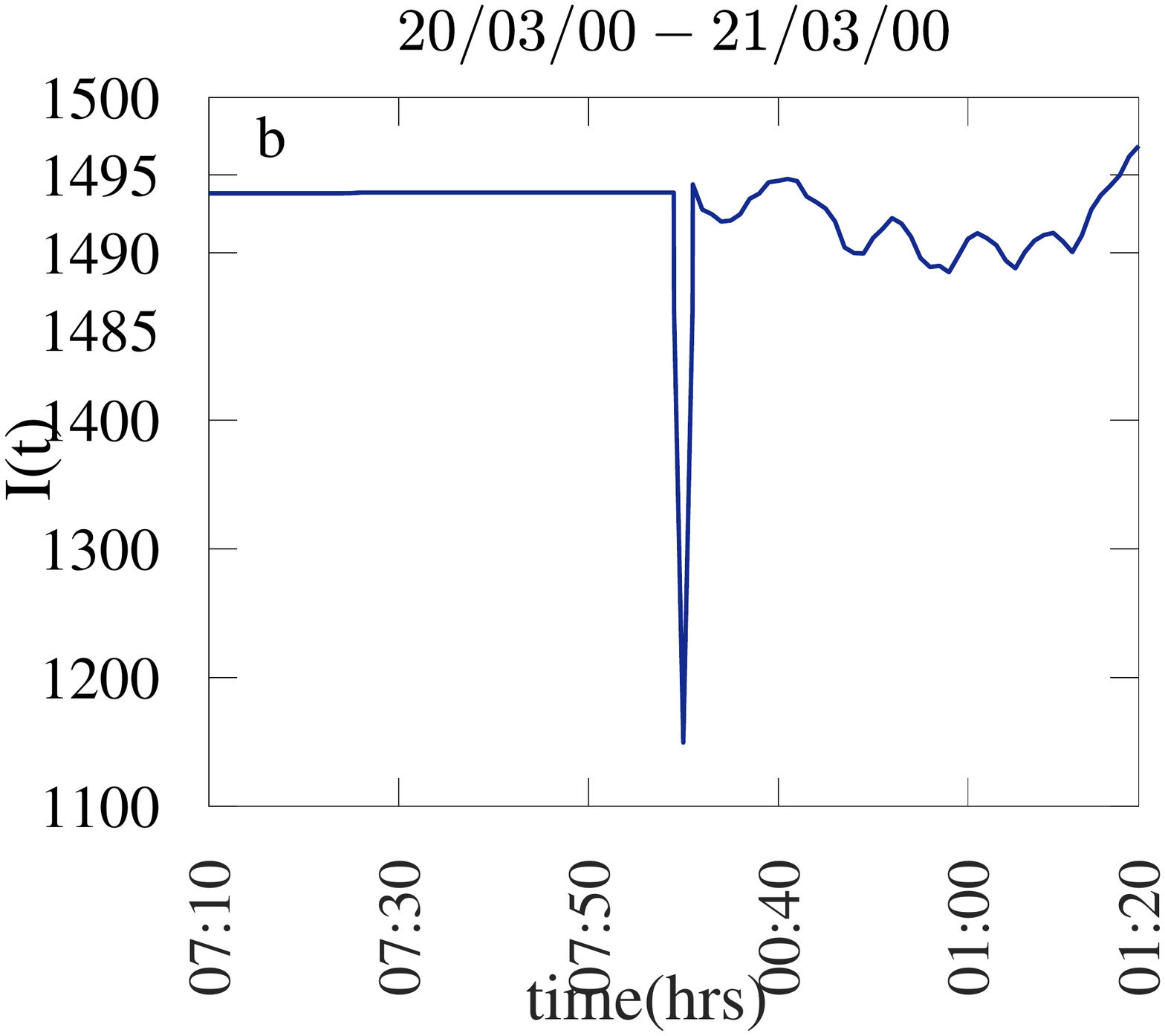}
	\includegraphics[scale=0.28,trim=0cm 5cm 0.0cm 4cm,clip=true]{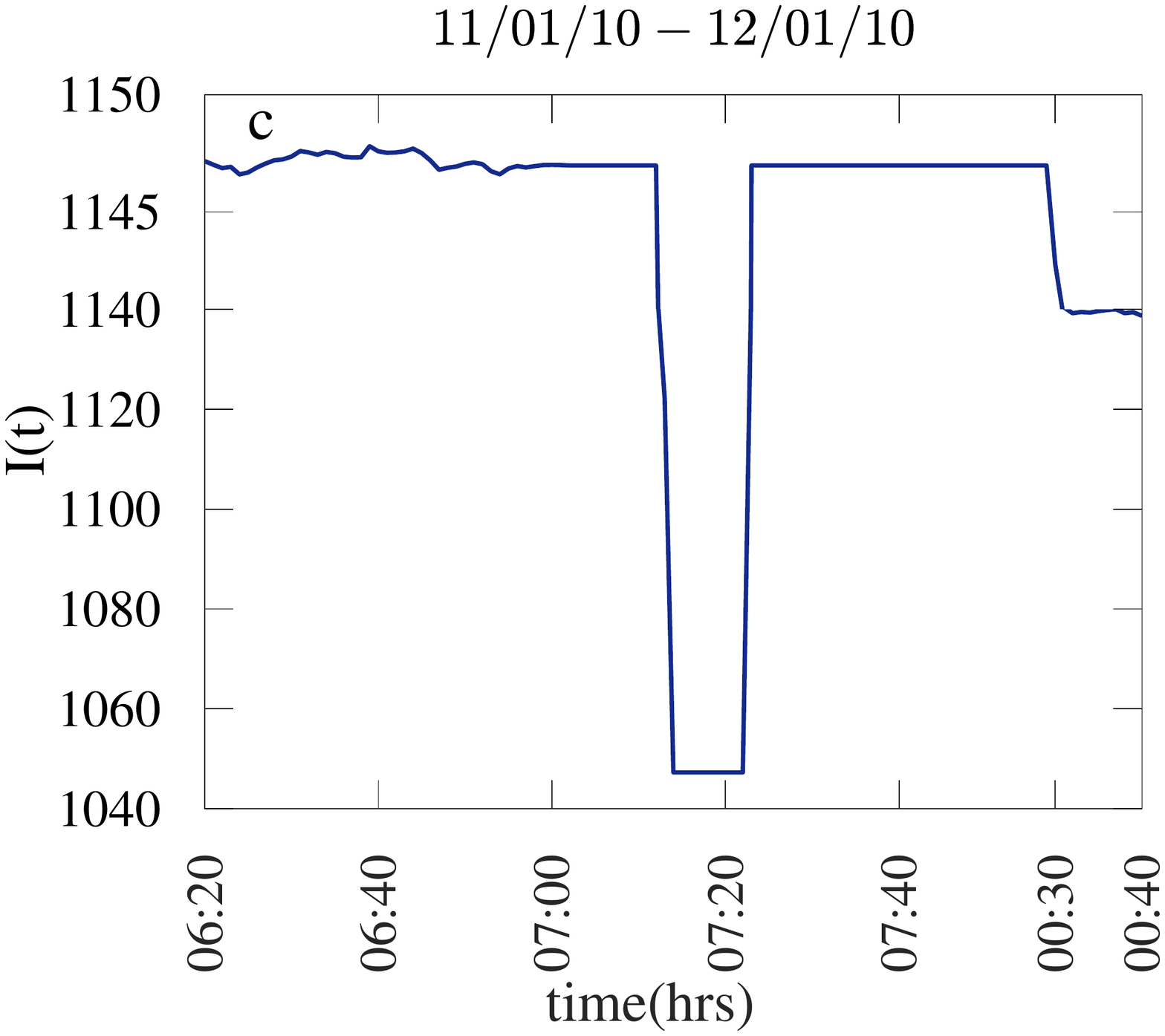}
	\caption{The three events that produce spurious jumps in S\&P500 data over the past 24 years are shown in subfigures (a) the peak of the  Asian financial crisis, (b) The imminent bankruptcy of many  Internet companies was announced,  and (c) The  Haiti earthquake effect.}
	\label{fig:Time_series_1}       
\end{figure}

The three artifacts removed from the S\&P500 data are: In 17/06/1997, at the peak of the Asian financial crisis \cite{Zen2020}. In 20/03/2000, when the imminent bankruptcy of many Internet companies was announced \cite{CNNfn2000}. The third artifact removed was the Haiti earthquake that occurred on 11/01/2010 \cite{ferreira2015earthquakes}. These three events introduce spurious jumps in the S\&P500, producing a considerable percentage change of the Index for $1$ to $8$ minutes only. Then, the abrupt Index values return to the original. The three artifacts are shown in Figure \ref{fig:Time_series_1}(a-c) in chronological order. \\

\section{Appendix B: Detrended Fluctuation Analysis (DFA)}
\label{AppendixB:DFA}
The Detrended Fluctuation Analysis method has become a widespread technique to determine the self-similarity of a time series based on statistical functions $F_{w}$ \citep{peng1994mosaic,kantelhardt2002multifractal}. 
The statistical functions $F_{w}$ [Eq.~(\ref{eq:DFA3})] are calculated based on central moments of the time series. For that, the detrended time series is divided into equal non-overlapping segments with a length of $s$. The $s$ value represents the time segment in which the detrended time series will be analyzed.

Then, the Hurst exponent is obtained from the power law given by $F_{w}$ vs $s$.\citep{peng1994mosaic,kantelhardt2002multifractal,kantelhardt2001detecting}. \par

\subsubsection{Description of the method of DFA}
We will proceed to indicate the steps to adopt this method to find the Hurst exponent of the detrended price return $x(t)$. First, we focus on calculating the statistical functions $F_{w}$, then to the calculation of Hurst exponent $H$ \citep{kantelhardt2002multifractal,gao2003principal,witt1998testing,kantelhardt2001detecting,horvatic2011detrended}.

\begin{description}[font=$\bullet$~\normalfont]
	\item [\textit{Step 1:}] The `profile' is the cumulative sum of the time series to analyze. For this case, our profile matches with the detrended price $I^{*}(t)$  obtained after applying Eq.~(\ref{eq:MA4}).
	\item [\textit{Step 2:}] The profile $I^{*}(t)$ is divided into non-overlapping segments $ \lfloor N_{s}=N/s \rfloor $ with the same length $s$. 
	\item [\textit{Step 3:}] The variance for each of the segments $v=1,2,3....N_{s}$ is obtained by applying the following equation,
	\begin{equation}\label{eq:MAk}
	F^{2}(v,s)=\frac{1}{s}{\sum_{i=1}^{s} (I^*[(v-1)s+i]-\overline{I^{*}}(v))^{2}},
	\end{equation}
	where, $\overline{I^{*}}(v)$ represents the mean of each segment of $I^{*}(t)$. 
	\item [\textit{Step 4:}] The statistical moments are obtained considering a range of values for the $\mathrm{w}^{\text{th}}$ orders.
	\begin{equation}\label{eq:DFA3}
	F_{\mathrm{w}}(s)=\left\lbrace \frac{1}{N_{s}}{\sum_{i=1}^{N{s}} [F^{2}(v,s)]^{\mathrm{w}/2}}\right\rbrace ^{1/\mathrm{w}}
	\end{equation}
\end{description}

\begin{figure*}[!htbp]
	\centering
	\includegraphics[scale=0.60,trim=3cm 2.0cm 1cm 11.0cm]{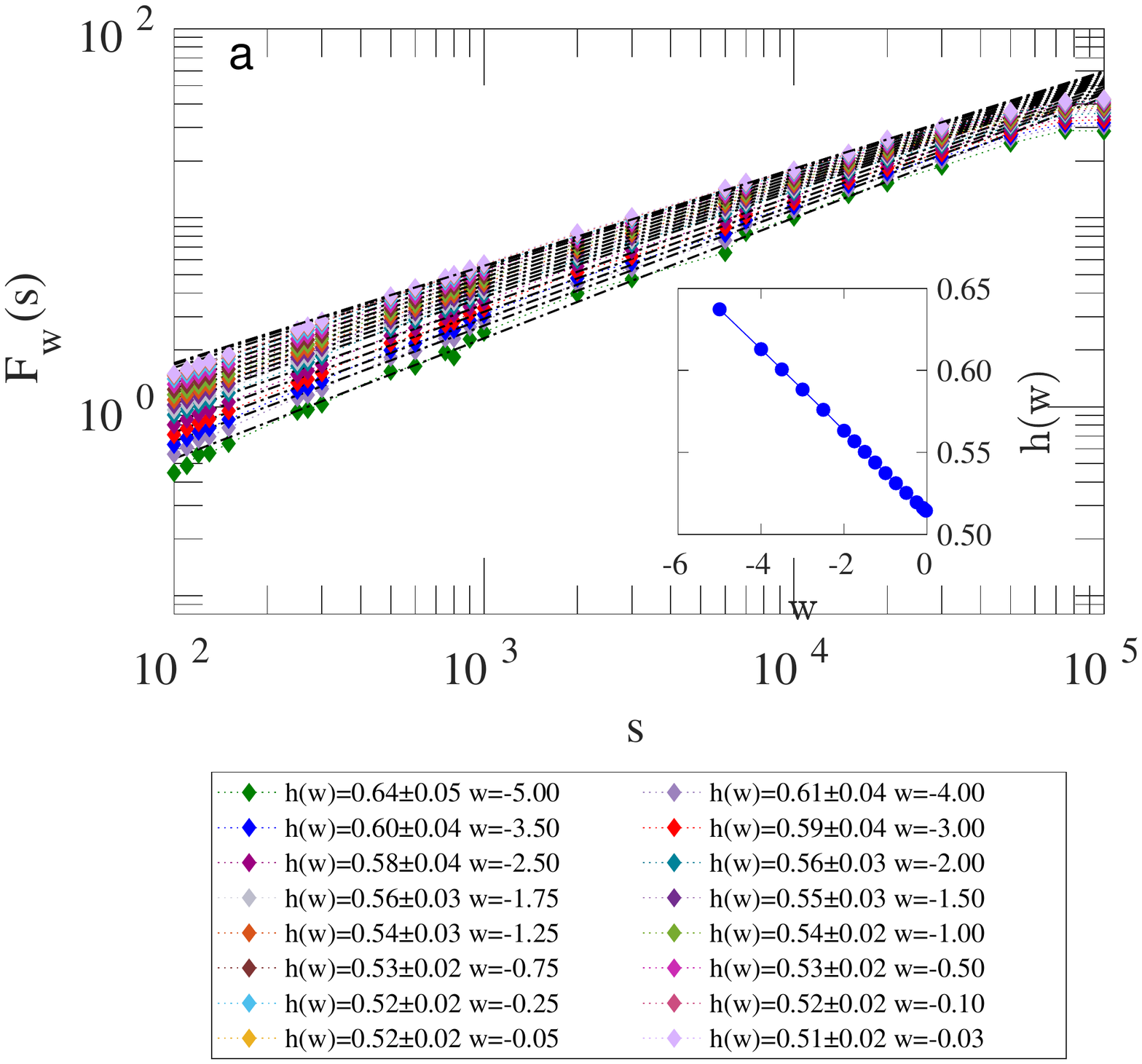}
	\includegraphics[scale=0.60,trim=3cm 2.0cm 1cm 9.0cm]{Fig_AppxB_Fw12monthsA}
	\caption{Calculation of the statistical function $F_w$ using Eq.~(\ref{eq:DFA3}). The function of $F_w$ vs $s$ display power laws $F_{w}(s) \sim s^{h(w)}$, where $h(w)$ depend on $w$. This feature demonstrates that the time series is a multifractal.}
	\label{fig:Moments}       
\end{figure*}

For detrended time series that present long-range correlations, the following power law is obtained \citep{kantelhardt2002multifractal,gao2003principal,kantelhardt2002characterization,rangarajan2000integrated}:
\begin{equation}\label{eq:Fw}
F_{\mathrm{w}}(s) \sim s^{h(\mathrm{w})}. 
\end{equation}
For monofractal time series $h(\mathrm{w})$ is independent from $\mathrm{w}$ value due to a constant scaling behaviour over all the segments ($F^{2}(v,s)$ in Eq.~(\ref{eq:MAk}) is cte.) \citep{maganini2018investigation,kantelhardt2002multifractal}, so $H=h(w)$. 
Figure \ref{fig:Moments} shows the results of DFA analysis, where a power law between $F_{\mathrm{w}}$ vs. $ s $ is observed.  However, the power laws depend on the  $\mathrm{w}^{\mathrm{th}}$ order. Consequently,  the Multifractal Detrending Fluctuation Analysis (MF-DFA) is applied to evaluate the dependence between the power laws and the order degree  $\mathrm{w}^{\mathrm{th}}$.

For a standard multifractal analysis (MF-DFA) the scaling exponent $\tau(\mathrm{w})$ needs to be obtained. This exponent is the power law defined between the $G_{\mathrm{w}}(s)$ vs $s$. The $G_{\mathrm{w}}(s)$ is a general statistical moment based on the following relationship:
\begin{equation}\label{eq:Ftau}
G_{\mathrm{w}}(s)={\sum_{v=1}^{N/s} \vert I^{*}(vs)-I^{*}((v-1)s)\vert^{\mathrm{w}}},
\end{equation}
where the term $I^{*}(vs)-I^{*}((v-1)s)$ is equal to the cumulative summation of the $x(t)$ within each segment $\nu$ of size $s$.
\begin{equation}\label{eq:MA2}
\sum_{t=(v-1)s+1}^{vs}x_{t}=I^{*}(vs)-I^{*}((v-1)s).
\end{equation}
In the standard multifractal formalism this sum is better known as the ``box probability" $p_{s}(v)$, and it is used to find hypothetical probability values on the original profile $I^{*}(t)$. 
\begin{equation}
p_{s}(v)=\sum_{t=(v-1)s+1}^{vs}x_{t}.
\label{Eq:ps}
\end{equation}\par
\par
Then, by replacing Eq.~(\ref{Eq:ps}) in Eq.~(\ref{eq:MA2}). The ``box probability" is defined in terms of the original profile, $ p_{s}(v)=I^{*}(vs)-I^{*}((v-1)s)$. After that, we replace this equivalence in Eq.~(\ref{eq:Ftau}), the following relation is obtained:
\begin{equation}\label{eq:Ftau1}
G_{\mathrm{w}}(s)={\sum_{v=1}^{N/s} \vert p_{s}(v)\vert^{\mathrm{w}}},
\end{equation}

\begin{figure*}[!htbp]
	\centering
	\includegraphics[scale=0.60,trim=3cm 2.0cm 1cm 11.0cm]{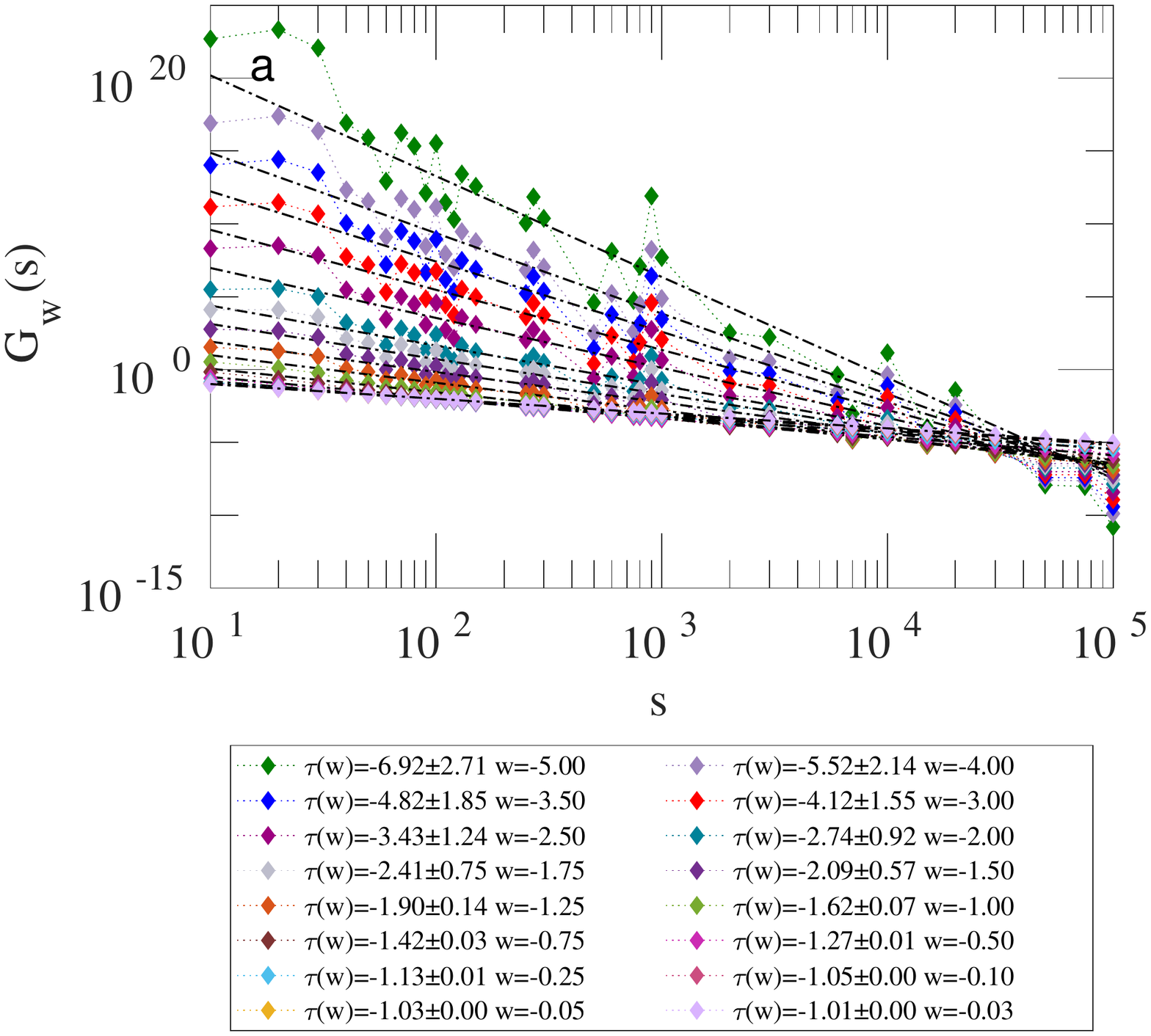}
	\includegraphics[scale=0.60,trim=3cm 2.0cm 1cm 9.0cm]{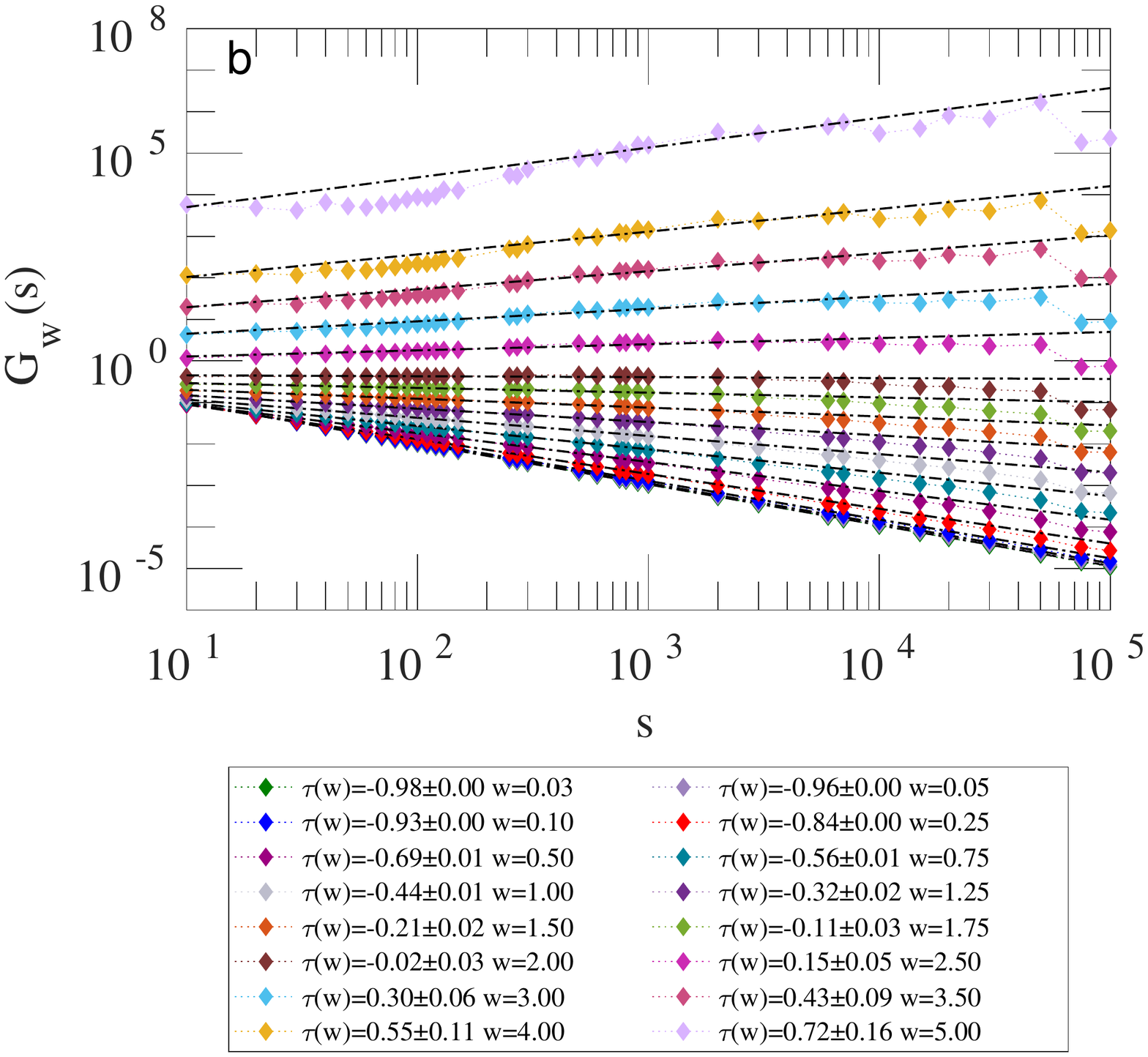}
	\caption{Calculation of the ``generalized statistical functions" $G_{\mathrm{w}}(s)$ vs $s$ for negative (a) and positive (b) values of order w. The ``generalized statistical functions" is calculated by applying Eq.~(\ref{eq:Ftau1}). }
	\label{fig:MomentsZq}       
\end{figure*}

where,
\begin{equation}\label{eq:F_tau_power_law}
G_{\mathrm{w}}(s) \sim s^{\tau(\mathrm{w})}.
\end{equation}

The $\tau(\mathrm{w})$ represents the classical multifractal scaling exponent, and is related with $h(\mathrm{w})$ for stationary and normalized time series \citep{maganini2018investigation,kantelhardt2002multifractal},
\begin{equation}\label{eq:tauq_H}
\tau(\mathrm{w})=\mathrm{w}h(\mathrm{w})-1.
\end{equation}
The Figure \ref{fig:tau_q}, shows a summary of the scaling exponents $h(\mathrm{w})$ and $\tau(\mathrm{w})$, respectively. The relationship between $h(\mathrm{w})$ and $\tau(\mathrm{w})$ is noticed, prooving that $x(t)$ is a multifractal stationarity time series.  The Figures \ref{fig:MomentsZq}a and \ref{fig:MomentsZq}b show the results of the generalized statistical function $G_{\mathrm{w}}$ for negative and positive values of $\mathrm{w}$ respectively. 

\end{document}